\documentclass[letterpaper,twocolumn,10pt]{article}
\usepackage{style/usenix}

\usepackage{algorithm}
\usepackage{amsmath}
\usepackage{amsthm}
\usepackage{amsfonts}
\usepackage{amssymb}

\usepackage{color}
\usepackage{enumitem}
\usepackage{fancyhdr}
\usepackage{float}
\usepackage{graphicx}
\usepackage{hyperref}

\usepackage{listings}

\usepackage[binary-units]{siunitx}
\usepackage{url}
\usepackage{xspace}
\usepackage{hyphenat}

\lstdefinelanguage{javascript}{
  keywords={typeof, new, true, false, catch, function, return, null, catch, switch, var, if, in, while, do, else, case, break},
  keywordstyle=\color{blue}\bfseries,
  ndkeywords={class, export, boolean, throw, implements, import, this},
  ndkeywordstyle=\color{black}\bfseries,
  identifierstyle=\color{black},
  sensitive=false,
  comment=[l]{//},
  morecomment=[s]{/*}{*/},
  commentstyle=\color{purple}\ttfamily,
  stringstyle=\color{red}\ttfamily,
  morestring=[b]',
  morestring=[b]"
}

\newcommand\myurl[2]{\url{#1}}

\newfloat{lstfloat}{htbp}{lop}
\floatname{lstfloat}{Listing}

\newenvironment{description*}%
  {\begin{description}%
    \setlength{\itemsep}{0pt}%
    \setlength{\parskip}{0pt}}%
  {\end{description}}
  
\newcommand{\anonword}[2]{%
\expandafter\newcommand\csname #1\endcsname{#2\xspace}%
}%

\begin{document}

\title{Restoring Uniqueness in MicroVM Snapshots}

\newcommand{\awsauthor}[2]{
{\rm #1}\\
\awsFull
}

\makeatletter
\iffalse
\author{
\awsauthor{Anonymous Authors}{@amazon.com}
}
\anonword{aws}{Anon Cloud Provider}
\anonword{amazon}{Anon}
\anonword{awsFull}{Anonymous}
\anonword{awslambda}{AnonFaaS}
\anonword{ebs}{AnonBS}
\anonword{ebsFull}{Anon Block Storage}
\anonword{fargate}{AnonPaaS}
\anonword{ectwo}{AnonIaaS}
\anonword{sthree}{AnonStore}
\anonword{sqs}{AnonQueue}
\anonword{kinesis}{AnonStream}
\anonword{dynamodb}{AnonDB}
\anonword{elasticache}{AnonCache}
\anonword{firecracker}{AnonVMM}
\anonword{kms}{Anoncloud Anonservice}
\anonword{enclaves}{Anoncloud TEE}
\else
\author{
\awsauthor{Marc Brooker}{@amazon.com}
\and \awsauthor{Adrian Costin Catangiu}{@amazon.com}
\and \awsauthor{Mike Danilov}{@amazon.com}
\and \awsauthor{Alexander Graf}{@amazon.com}
\and \awsauthor{Colm MacCarthaigh}{@amazon.com}
\and \awsauthor{Andrei Sandu}{@amazon.com}
}
\anonword{aws}{AWS}
\anonword{amazon}{Amazon}
\anonword{awsFull}{Amazon Web Services}
\anonword{awslambda}{Lambda}
\anonword{ebs}{EBS}
\anonword{ebsFull}{Elastic Block Storage}
\anonword{fargate}{Fargate}
\anonword{ectwo}{AWS EC2}
\anonword{sthree}{S3}
\anonword{sqs}{SQS}
\anonword{kinesis}{Kinesis}
\anonword{dynamodb}{DynamoDB}
\anonword{elasticache}{Elasticache}
\anonword{firecracker}{Firecracker}
\anonword{kms}{AWS Key Management Service}
\anonword{enclaves}{AWS Nitro Enclaves}
\fi
\makeatother

\markboth{DRAFT. Amazon Confidential. Not for distribution or publication.}%
{DRAFT. Amazon Confidential. Not for distribution or publication.}

\maketitle

\section*{Abstract}
Code initialization---the step of loading code, executing \emph{static} code, filling caches, and forming re-used connections---tends to dominate cold-start time in serverless compute systems such as \aws \awslambda. Post-initialization memory snapshots, cloned and restored on start, have emerged as a viable solution to this problem, with incremental snapshot and fast restore support in VMMs like \firecracker.

Saving memory introduces the challenge of managing high-value memory contents, such as cryptographic secrets. Cloning introduces the challenge of restoring the uniqueness of the VMs, to allow them to do unique things like generate UUIDs, secrets, and nonces. This paper examines solutions to these problems in the \emph{every microsecond counts} context of serverless cold-start, and discusses the state-of-the-art of available solutions. We present two new interfaces aimed at solving this problem---\texttt{MADV\_WIPEONSUSPEND} and VmGenId---and compare them to alternative solutions.
\section{Introduction}\label{sec:introduction}

When \aws \awslambda receives a request to invoke a serverless function, the system attempts to match the request to an already-running microVM containing a copy of the function code. At scale, this \emph{warm start} succeeds with high probability--both because of the averaging effects of scale, and because serverless systems attempt to predict the number of incoming requests using techniques such as reinforcement learning~\cite{Balaji2020}, LSTMs~\cite{gunasekaran2020}, and ARIMA models~\cite{shahrad2020}. When warm start does not succeed, the system falls back on \emph{cold start}, which involves starting a \firecracker~\cite{agache2020} microVM (or taking one from a pool), loading the function code, and initializing the code to get it ready to handle the request. The last step of the the cold start, initialization, is typically the one that takes longest. Du et al~\cite{Du2020} found this initialization step to take between 25\% and 95\% of total start-up time, mirroring our own experiences.  As an additional challenge for cloud providers, this initialization step is almost entirely within the customer's control, leaving little opportunity to optimize or improve it on behalf of the customer (beyond simply allocating additional resources to its execution).

\begin{figure}[!t]
\centering
\includegraphics[width=0.5\columnwidth]{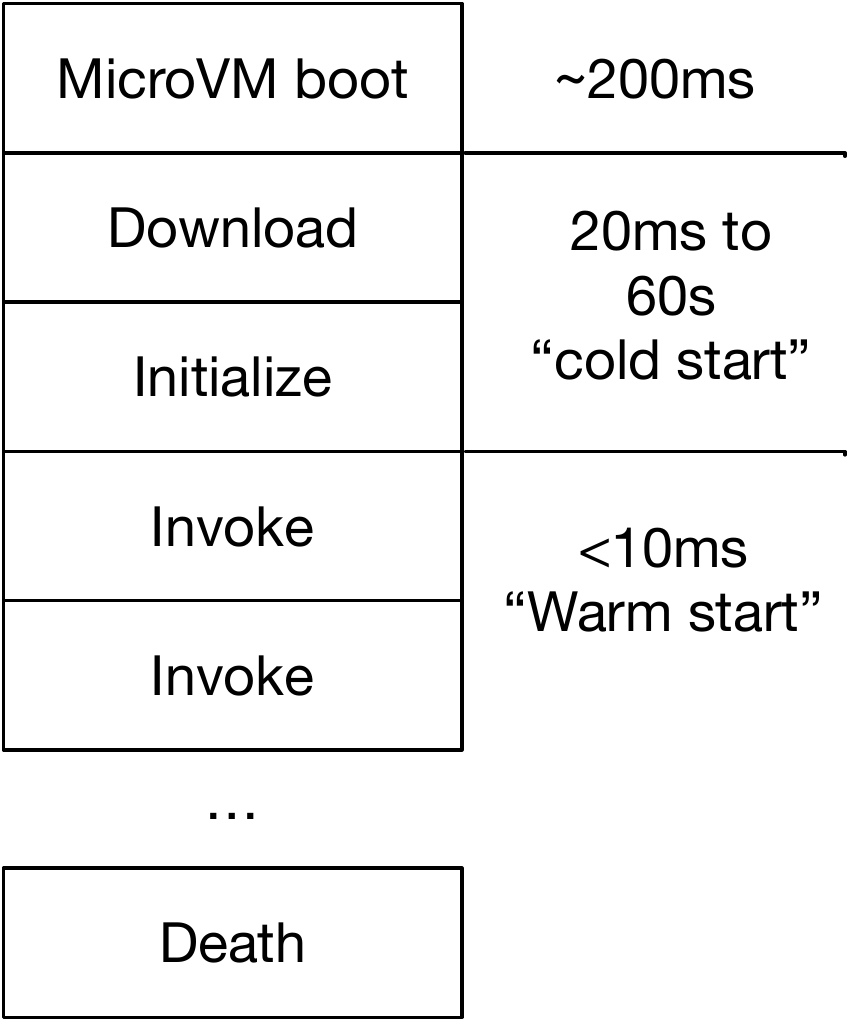}
\caption{Time line of the life cycle of a microVM in \aws \awslambda}
\label{fig:timeline}
\end{figure}

Figure \ref{fig:timeline} shows a typical time line of the life cycle of a microVM in \awslambda. The microVM boot, booting a minimal Linux kernel in \firecracker, with a minimal set of userspace daemons, and network setup, takes approximately 200ms. This is hidden from the customer through the use of a small pool of pre-booted microVMs. The cold start portion, including downloading the function code into the microVM and initialization, frequently takes up to 60s but can be as little as 20ms for small static binaries. The download is typically short: at 25Gb/s, downloading \awslambda's maximum package size of 250MB takes only 80ms. Finally, each invoke experiences a warm start delay caused by routing, typically less than 10ms. The same microVM is then re-used, for future invokes of the same version of the same function, to amortize the cold start costs.

The time and resources taken by initialization depend strongly on the customer's choice of language and framework, and on additional work the customer has chosen to do at process start time. Of the popular languages, Java typically exhibits the longest cold-start times. This is driven by three main factors: the start-up time of the JVM itself, the decompression and loading of class code, and the execution of \emph{static} code in the loaded classes. It is typically the last of these which drives long cold-start times, because Java allows for the execution of effectively arbitrary code at static initialization time. While Java code initialization and static execution is multithreaded, the specification~\cite{jls11} limits parallelism by defining a strict recursive order on class initialization. This problem is not only limited to Java. The same outcome is common, but less ubiquitous, in workloads written in Javascript, Python, and C\#. Workloads written in native languages like C, C++, Rust and Go tend to initialize quickly, largely because it's a less-common pattern to do extensive work at startup in these languages.

Post-initialization snapshots have been proposed as a solution to this initialization time problem~\cite{Du2020, Cadden2020}, to the general problem of cloud scaling~\cite{bryant2011}, to fault-tolerance~\cite{lorch2015}, and even for accelerating hypervisor fuzzing~\cite{schumilo2021}. Simply, initialization is complete, the virtual machine is snapshotted, and when a cold start is needed the snapshot is cloned and restored. This approach is attractive, because microVM restore can be made very fast. In our own experiments, \firecracker snapshot restore can take as little as 4ms. Catalyzer~\cite{Du2020} reports latencies reliably below 10ms for a broad range of application types.

\begin{figure}[!t]
\centering
\includegraphics[width=\columnwidth]{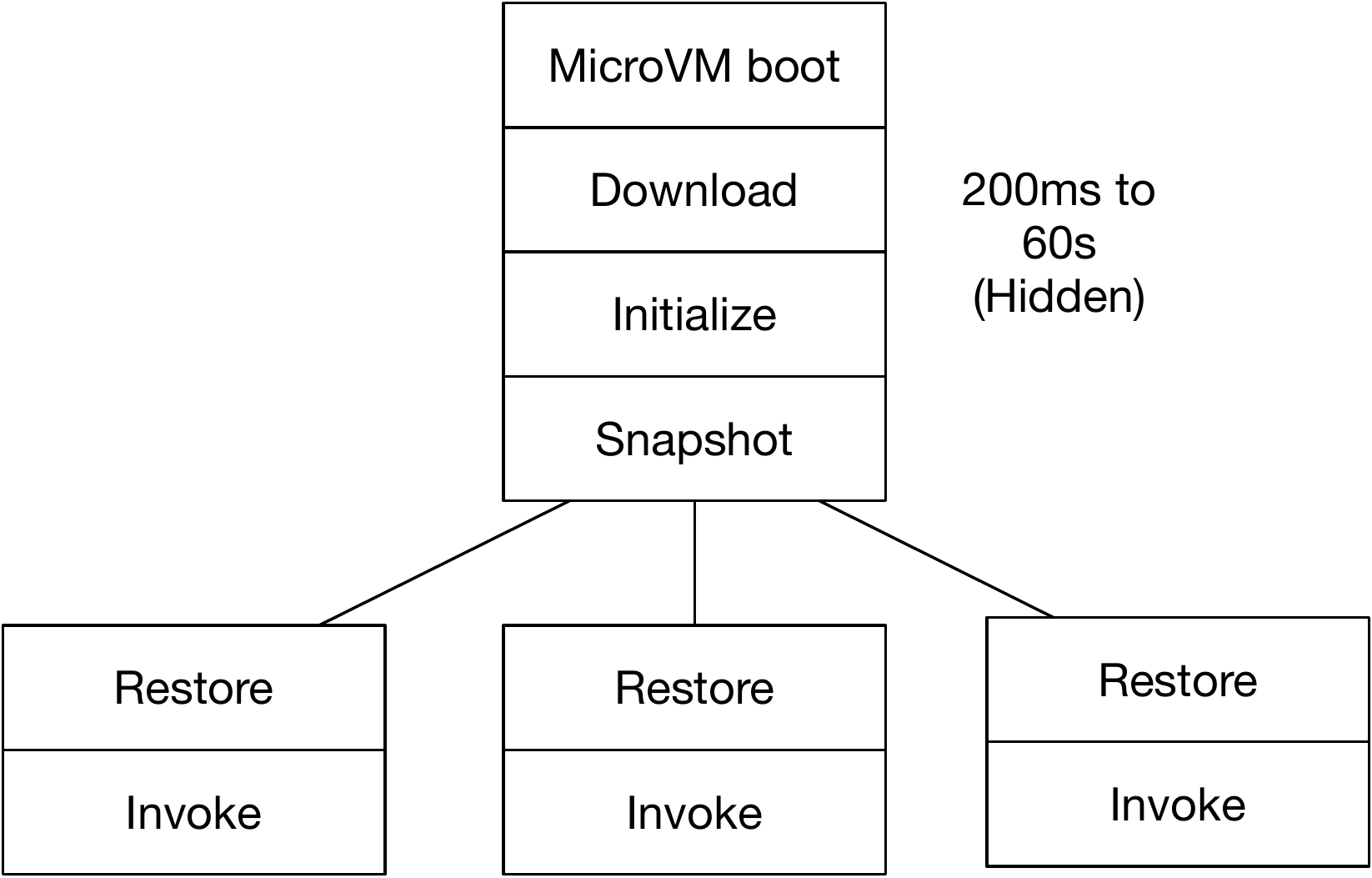}
\caption{Time line of the life cycle of a microVM in \aws \awslambda with snapshot and clone}
\label{fig:timelineclone}
\end{figure}

Figure \ref{fig:timelineclone} shows the timeline for a microVM with snapshot and clone. The entire boot and cold start portion can be hidden from the customer by performing it once at function creation time. The blocking portion then requires only a snapshot restore, a much faster operation for which the entire latency is under the control of \aws.

\subsection{Challenges}
Snapshotting and clone-on-restore introduce several system-level challenges. First, by their nature snapshots turn data in RAM into data on storage. This data can include cryptographic secrets and other high-value tokens which the programmer wished to control carefully, irrespective of the security properties of the snapshot storage and distribution layer. Second, cloning duplicates memory state, which is a challenge for applications which want to generate unique data such as request IDs, UUIDs, and cryptographic nonces. Third, cloning is not compatible with existing session-layer network protocols such as TCP and TLS, which assume that the client is a single unique entity with a single identity. Re-establishing these connections can add significantly to restore latency~\cite{hunhoff2020, hunhoff2020poster}, and can even cause correctness issues in some protocols (a type of replay attack). This paper focuses on the second challenge---uniqueness---but some of the solutions we analyze also help address the challenges posed by high-value secrets and network protocols.

\section{The Value of Uniqueness}\label{sec:uniqueness}

Most modern distributed systems depend on the ability of nodes to generate unique or random values. RFC4122~\cite{rfc4122} version 4 UUIDs are widely used as unique database keys, and as request IDs used for log correlation and distributed tracing. In this context, duplicate IDs will lead to correctness issues, logical data corruption, incorrect traces, and confusing or misleading logs. UUIDs are also used as tokens for API idempotency, where multiple calls with the same token behave as a single call. Here, duplicate IDs can lead to availability or correctness issues. Many common distributed systems protocols, including consensus protocols like Paxos~\cite{Lamport2001}, and ordering protocols like vector clocks, rely on the fact that participants can uniquely identify themselves. Cloning without changing identity could be seen as an unintentional Sybil~\cite{douceur2002} attack, which most distributed systems protocols don't tolerate.

Jitter, the intentional addition of pseudorandomness to values like retry intervals, sampling intervals, timeouts and resource limits are widely used to improve the resilience of distributed systems. For example, adding jitter to the retry intervals used by clients of an overloaded system reduce the clustering of retries in time, helping systems recover faster from transient overload. Adding jitter to periodic jobs, like \texttt{cron}, helps avoid seasonalities like start-of-day and top-of-hour spikes. Systems that aren't unique will tend to cluster, and clustering leads to correlated load, which leads to lower system utilization.

Cryptography is the most critical application of unique data. Any predictability in the data used to generate cryptographic keys---whether long-lived keys for applications like storage encryption or ephemeral keys for protocols like TLS---fundamentally compromise the confidentiality and authentication properties offered by cryptography. A potentially less-obvious vulnerability relates to initialization vectors (IVs) and nonces used in common block cipher modes like counter mode (CTR) and Galois counter mode (GCM). Using the same IV with the same key multiple times breaks the security of these modes. In many protocols, the IV is translated in the clear, making it easy for an attacker to tell if an IV has been re-used. Dworkin in NIST SP800-38D~\cite{dworkin2007} says:

\begin{quote}
The probability that the authenticated encryption function ever will be invoked with the same IV and the same key on two (or more) distinct sets of input data shall be no greater than $2^{-32}$.

\ldots

In practice, this requirement is almost as important as the secrecy of the key.
\end{quote}

These properties---the need for unique IVs and keys with both low predictability and very small probability duplication---require that even cloned MicroVMs have access to high-quality entropy, and the means to use it. In some cases, such as AES-GCM, these values need only be unique and not random, so access to a unique per-VM identifier is sufficient. Solutions to the entropy problem are well-understood, and widely deployed in virtualized environments. Typically, entropy is injected into the guest, where it can be accessed using devices (like \texttt{/dev/urandom}) or system calls (like \texttt{getrandom}). Hardware RNGs, such as Intel's \texttt{RDSEED} and \texttt{RDRAND}~\cite{inteldev2013} (itself the output of a DRBG seeded by a hardware RNG), can also provide high-quality entropy inside cloned MicroVMs. We aren't the first to notice these kinds of cryptographic vulnerabilities. For example, Ristenpart and Yilek~\cite{ristenpart2010} describe successful attacks against TLS 1.0 relying on the-reuse of VM snapshots.

Many applications don't obtain their entropy directly from the kernel or hardware, and instead deploy userspace pseudo-random number generators (PRNGs) or deterministic random bit generators (DRBGs). Widely-deployed cryptographically-secure PRNGs include NIST's CTR\_DRBG~\cite{barker2015}, Java's \texttt{SecureRandom}, and Javascript's \texttt{getRandomValues}. Most of these PRNGs are deterministic algorithms that get their seeds from the kernel or hardware RNGs, either at startup or periodically, and \emph{stretch} them into an unpredictable sequence of random bits.

This presents a challenge to the serverless compute provider: even if they follow best practices such as reseeding kernel randomness, and customers follow best practices like using a cryptographically secure PRNG (CSPRNG), the combination is not secure. For platforms like \aws \awslambda which allow customers to run arbitrary code, even arbitrary binaries, addressing this challenge is not trivial.

\section{Userspace Interfaces for Uniqueness}\label{sec:interfaces}

Solving the uniqueness problem strongly enough for cryptographic purposes requires a mechanism which can deterministically reseed userspace PRNGs with new entropy at restore time.
This mechanism must also support the high-throughput and low-latency use-cases that led programmers to pick a userspace PRNG in the first place (so simply reverting to \texttt{getrandom} is not acceptable); be usable by both application code and libraries; allow transparent retrofitting behind existing popular PRNG interfaces without changing application code; it must be efficient, especially on restore; and be simple enough for wide adoption.

Efficiency is a particularly important concern. The default PRNGs in many languages, including C, Java, and Python, are not cryptographically secure, a choice typically driven by performance. This frequently leads to programmers using these weak random number generators for cryptographic purposes. Over 350 CVEs have been allocated in 2020 alone for security issues introduced by this class of bugs. Reducing the efficiency of CSPRNGs further would increase adoption of non-CS PRNGs, leading to more bugs and insecure cryptography implementations.

\subsection{Fork and \texttt{MADV\_WIPEONFORK}}
A similar problem with userspace PRNGs is introduced by \texttt{fork}, as processes memory is duplicated in both the parent and child processes. A new \texttt{madvise} flag, \texttt{MADV\_WIPEONFORK}, was introduced in Linux 4.14 in 2017.
Memory pages marked \texttt{MADV\_WIPEONFORK} are set to all zeros in the child process after the call to \texttt{fork}, \texttt{clone} and related calls.
PRNGs which mark their internal state this way, or put a guard variable in a page marked this way, can deterministically detect they have been forked, and reseed from kernel or hardware randomness.
This does not require additional system calls, and adds only the overhead of a single memory access per call to the random number generation library.
BSDs support a similar approach, using the \texttt{MAP\_INHERIT\_ZERO} flag to the \texttt{minherit} system call.

The same interface is useful in other ways too. A process which handles cryptographic keys or other high-value material can mark them as \texttt{MADV\_WIPEONFORK} to ensure that they are not needlessly copied into child processes. It can also be combined with \texttt{mlock} and \texttt{MADV\_DONTDUMP} to ensure that secrets aren't saved onto disk.

\texttt{MADV\_WIPEONFORK} is superficially similar to \texttt{pthread\_atfork}, but is more robust because it is not thread-specific, and works across all ways to clone a process. \texttt{pthread\_atfork} can be bypassed, intentionally or unintentionally, by using the \texttt{clone} syscall directly, and by some other interfaces like \texttt{posix\_spawn}. This makes \texttt{pthread\_atfork} insufficiently robust for use in widely-used libraries.

\subsection{Suspend and \texttt{MADV\_WIPEONSUSPEND}}
By analogy to \texttt{MADV\_WIPEONFORK}, we have contributed a new \texttt{MADV\_WIPEONSUSPEND} flag to the Linux kernel\footnote{Our contributions here focus on Linux because we use it as the guest OS in \aws \awslambda's microVMs, but are widely applicable to other guest operating systems, unikernels, and library OSs}, which marks pages to be wiped when a VM is suspended (a precursor to snapshotting). We expect that PRNGs and cryptographic libraries mark their state or guard variables as both \texttt{MADV\_WIPEONFORK} and \texttt{MADV\_WIPEONSUSPEND} and use the same reseeding logic to handle both the fork and VM clone cases.

Wiping on suspend, rather than restore, has two benefits. One is performance: in our serverless use-case, restore is on the latency-critical cold-start path, so handling memory wiping at suspend time optimizes restore latency. Second is security: applications which have high-value secrets they don't want to include in the snapshot, or intentionally want to retrieve directly from a key management service or hardware security module (HSM) on restore, can mark these secrets to be excluded from the snapshot.

\texttt{MADV\_WIPEONSUSPEND} has the benefits of being simple, having very little performance overhead, and being easy to use both in libraries and application code. The performance overhead of \texttt{MADV\_WIPEONFORK} and \texttt{MADV\_WIPEONSUSPEND} is small, about 600ns per random generation. In an implementation of NIST's CTR\_DRBG, performing the necessary checks reduced per-core throughput by just TODO\%.

\subsection{System Generation ID}
The Windows Server 2012 version of the Hyper-V hypervisor introduced the VM Generation ID (VmGenId). The guest OS detects that significant changes have occurred, including cloning and restore, by a change in the VmGenId. Microsoft's VmGenId implementation~\cite{VmGenId2012} exposes a 128-bit unique number, a UUID, in a memory location that is discovered using a special ACPI device name. While interacting with the low-level implementation is not convenient for applications and libraries, and they are unlikely to have permissions to do so, the VmGenId is used in the Windows Cryptography API to avoid problems with cloning. Other hypervisors, including VMWare and QEMU, have since added support for VmGenID.

We have contributed support for a similar mechanism to the Linux kernel, called System Generation ID (SysGenId). The Linux version implements a device driver which exposes a read-only device \texttt{/dev/SysGenId} to userspace, which contains a monotonically increasing generation counter. Libraries and applications are expected to \texttt{open()} the device, and then call \texttt{read} which blocks until the SysGenId changes. Following an update, \texttt{read()} calls no longer block until the application acknowledges the new VmGenId by \texttt{write}ing it back to the device. Non-blocking \texttt{read} calls return \texttt{EWOULDBLOCK} when their is no new SysGenId available. Alternatively, libraries can \texttt{mmap} the device to get a single shared page which contains the latest VmGenId at offset 0.

Linux SysGenId also supports a notification mechanism exposed as two \texttt{ioctls} on the device. \texttt{SysGenId\_GET\_OUTDATED\_WATCHERS} immediately returns the number of file handles that were open during the last SysGenId change but have not yet acknowledged the new id. \texttt{SysGenId\_WAIT\_WATCHERS} blocks until there are no open file handles on the device which haven't acknowledged the new id. These two interfaces are intended for serverless and container control planes, which want to confirm that all application code has detected and reacted to the new SysGenId before sending an invoke to the newly-restored sandbox. This notification mechanism, unlike signals, allows SysGenId to be used inside libraries without requiring changes to application code.

The Linux SysGenId implementation also supports a VmGenId, as defined by Microsoft. SysGenId is the frontend driver exposing the 32-bit generation ID, which depends on a backend driver to detect when the system's identity changes. VmGenId is one such backend. when the VmGenId changes, the SysGenId is increased. Other backends to SysGenId can be defined, making it more general, and allowing support for cases like userspace checkpoint restore (e.g. CRIU~\cite{Criu2021}), and Linux containers. In this sense, SysGenId is a generalization of VmGenId.

\subsection{Comparing Solutions}
The three solutions \texttt{MADV\_WIPEONSUSPEND}, Microsoft's VmGenId, and Linux SysGenId) to handling uniqueness during VM cloning are compared in Table \ref{table:comparison}. In \texttt{MADV\_WIPEONSUSPEND}'s favor is its compatibility with \texttt{MADV\_WIPEONFORK}, its additional use for excluding secrets from snapshots, and the fact that it can be used in containers and sandboxes that don't have the ability to open or mmap files. The popularity of container technology, and sandboxing approaches like \texttt{seccomp-bpf} make this property especially interesting. Many sandboxes and containers want to prevent processes from opening files and devices entirely. \texttt{open}, \texttt{read} and \texttt{write} give a potential attacker a lot more power than just \texttt{madvise}.

In SysGenId's favor is its flexibility, including the ability to monitor for when processes and libraries have fully caught up to the latest version of a container. On Linux, both solutions have little memory overhead (involving reading from a special page), and can be used by non-root processes. This same flexibility comes with some risk. By enabling libraries and applications to do blocking work on the restore path (while the system blocks on \texttt{SYSGENID\_WAIT\_WATCHERS}), they can re-introduce a form of cold-start latency. Application and library authors must be encouraged to use SysGenId for lightweight purposes only.

\begin{table}[ht]
\begin{tabular}{ |l|lll| }
\hline
 & MADV & VmGenId & SysGenId \\
\hline
Mechanism & Guard Page & UUID & Inc. Id \\
Works for fork & Yes & No & No \\
Secret hiding & Yes & No & No \\
In-memory & Yes & Yes & Yes \\
Notification & No & No & Yes \\
Non-root & Yes & No & Yes \\
Min-privilege & Yes & No & No \\
Entropy & No & Yes & No \\
Containers & No & No & Yes \\
\hline
\end{tabular}
\caption{Comparison of features of \texttt{MADV\_WIPEONSUSPEND}, Microsoft's VmGenId, and Linux SysGenId}
\label{table:comparison}
\end{table}

In Microsoft's solution's favor is its use of a high-quality UUID VmGenId, which can be used directly as a node identifier for distributed protocols, for logging, and tracing. The UUID, which Microsoft~\cite{VmGenId2012} describe as \emph{"a 128-bit, cryptographically random integer value identifier"} can also be used directly as a high-entropy seed for PRNGs, removing the need to separately reseed from the kernel or hardware. It can also be used directly for the deterministic construction of IVs (see NIST SP800-38D, section 8.2.1). Microsoft's VmGenId has existed for longer than either of the Linux solutions, and is built into the Windows Cryptography API, meaning that it already has extensive userspace support.

\subsection{When is a VM the same VM?}
While \texttt{MADV\_WIPEONSUSPEND} has fairly obvious semantics, SysGenId raises the question of under which circumstances the ID should be changed. Microsoft change their VmGenId on restore, copy, clone, recovery from backup, and (in some cases) on failover. The VmGenId does not change on reboot, pause, resume, live migration, or even on host reboot. While most of these seem like reasonable decisions, it is not clear that they are a good match for a serverless environment.

Consider the case of restoring from a snapshot without cloning. The resulting microVM is, for practical purposes, the same microVM as the one that was snapshotted, as long as the system can reason that it can never be clones. This means that SysGenId is just that---a generation id---and not the id of the microVM itself. Figuring out the right ID for a microVM is a separate system-level concern.

\subsection{Alternative approaches}
One alternative to approaching this problem at the system level is to approach it at the level of cryptographic algorithms. Rogaway and Shrimpton proposed~\cite{rogaway2007, rogaway2007siv} authenticated encryption modes that do not depend on a random nonce for their strongest security properties. Gueron et al proposed~\cite{gueron2015} and standardized~\cite{rfc8452} AES-GCM-SIV provides a nonce misuse-resistant mode, which keeps it's confidentiality and authentication properties in case of IV re-use (exposing only message equality) with only small performance overhead. Ristenpart~\cite{ristenpart2010} proposed changes to the TLS protocol to resist nonce-reuse. Despite these advances, and many others, many of the cryptographic protocols and modes widely deployed in production remain vulnerable to nonce reuse. We do not expect this situation to change quickly.

Another alternative is to move towards deprecating userspace CSPRNGs, and encourage implementors to use random numbers provided by the kernel or hardware. Discussing the merits of this approach is beyond the scope of this paper, but it is clear than making such a change is not possible without changing the performance of many libraries and applications, and changing many pieces of software to adopt new RNG interfaces. A system-level approach allows us to approach this change incrementally, which is essential given the size of the installed base and the goal of serverless platforms to work with existing application software.

A similarly major architectural change is to build applications which rely on cryptographic services provided by hardware security modules (HSMs), key management services (such as \kms), trusted execution environments (TEEs), and secure enclaves (such as \enclaves). In many cases this would be a good choice, but again isn't always applicable due to application-specific requirements, data plane performance requirements, and large bodies of existing code.

\subsection{Atomicity and TOCTOU}
Both the \texttt{MADV\_WIPEONSUSPEND} and SysGenId options provide sufficient mechanism for a RNG library to ensure that it does not generate duplicate random numbers in multiple cloned VMs\footnote{TLA+ specifications demonstrating this property are available at http://redacted.example.com/}. In the SysGenId case, this requires that the SysGenId update happens before (or atomically with) the restore operation. It is up to the VMM and guest kernel to ensure that this ordering is not violated.

Serverless systems (like \aws \awslambda) explicitly track manage the requests in-flight in a running VM, and ensure that VMs cannot perform any actions outside the context of handling a request. These systems can therefore fully quiesce a VM and ensure that it does not handle any requests until the notification process is complete. This is sufficient to ensure that random numbers are never re-used across multiple requests, provided that application code does not cache values between requests.

In general, however, none of these mechanisms are sufficient to strongly prevent re-use, due to \emph{time of creation time of use} (TOCTOU) issues. In any program which does not atomically generate, use, and discard the generated value, a clone operation between generation and use, or between use and discard, can cause re-use. Atomicitiy of such complex operations is not generally supported. Some external fence provided by the environment is needed to ensure that reseeding is complete before connections are made or used, storage is modified, or the system has any other side-effect. Load balancer and container orchestrator health checks are likely good places to tie in this logic.

\section{Benchmarks}\label{sec:benchmarks}

In order to understand the steady-state and restore-time performance impact of these mechanisms, we ran a number of benchmarks on typical x86 (EC2 m5.12xlarge) server. We first considered the impact of reseeding: getting entropy from the kernel (\texttt{/dev/urandom}) or hardware (\texttt{RDRAND} or equivalent), and passing it into the userspace PRNG (using openssl 1.1's \texttt{RAND\_seed}).

This basic reseeding is, as expected, very fast. Reseeding 32 bytes from \texttt{/dev/urandom} (including \texttt{open}, \texttt{read}, and \texttt{close}) took a mean of $11\mu s$, with insignificant deviation. Reseeding 32 bytes from \texttt{RDRAND} or \texttt{RDSEED} took $0.6\mu s$ per run with insignificant deviation (consistent with Intel's advertised performance for RDRAND and RDSEED~\cite{intelrng2018}).

Checking the guard page (either a page mapped with \texttt{MADV\_WIPEONSUSPEND} or the \texttt{mmap}ed VmGenId page) is, again, very fast. For the simple case of generating a sequential nonce (just a 128-bit increment operation, in other words), enabling the check reduces throughput by 13x. Doing any kind of meaningful work inside the check, on the other hand, causes its cost to amortize to near-zero. For example, running OpenSSL's default \texttt{md\_rand} generator with or without the guard page check showed no statistically significant difference. Absolute differences will vary from platform to platform, but we expect these relative results to be durable across platforms.
\section{Conclusion}\label{sec:conclusion}
Snapshot, clone and restore are useful and powerful primitives for reduced cold start times in serverless compute platforms. Using these primitives introduces several challenges: breaking common network protocols, moving in-memory state onto disk, and losing the ability for short-lived clones to make unique decisions. We described two new kernel interfaces that we proposed for tackling this problem in context of Linux and \aws \awslambda, one of which has been contributed to the mainline Linux kernel. This mechanism, a Linux flavor of VmGenId, provides a flexible way to approach these problems, with little performance overhead, relatively easy adoption, and wide applicability.

\bibliographystyle{plainurl}
\bibliography{paper}
\end{document}